\documentclass[
  twocolumn,
  dvipsnames]        %.. eg, OliveGreen 
  {aastex631}

  \usepackage{amssymb,amsmath,soul}
  \usepackage[capitalise]{cleveref}
  \newcommand{\naive} {$\hat{S}_2^\text{naive}(\tau)$}
  \newcommand{\lint }{$\hat{S}_2^\text{LINT}(\tau)$}
  \newcommand{\corr} {$\hat{S}_2^\text{corr}(\tau)$}

 \renewcommand{\d} {\mathrm{d}}
   \newcommand{\e} {\mathrm{e}}           %.. for exp

\begin{document}

\title{De-Biasing Structure Function Estimates From Sparse Time Series of the Solar Wind: A Data-Driven Approach
}

\correspondingauthor{Daniel Wrench}
\email{daniel.wrench@vuw.ac.nz}

\author[0000-0002-7463-3818]{Daniel Wrench}
\affiliation{Victoria University of Wellington \\ 
Kelburn, Wellington 6012, New Zealand}

\author[0000-0003-0602-8381]{Tulasi N. Parashar}
\affiliation{Victoria University of Wellington \\ 
Kelburn, Wellington 6012, New Zealand}

% \author[0000-0001-5840-8760]{Marcus Frean}
% \affiliation{Victoria University of Wellington \\ 
% Kelburn, Wellington 6012, New Zealand}

\begin{abstract}
Structure functions (SFs), which quantify the moments of increments of a stochastic process, are essential complementary statistics to power spectra for analyzing the self-similar behavior of a time series. However, many real-world datasets, such as those from spacecraft monitoring the solar wind, contain gaps, which inevitably corrupt the statistics. The nature of this corruption for SFs remains poorly understood\textemdash indeed, often overlooked. In this study, we simulate gaps in a large set of Parker Solar Probe magnetic field intervals to characterize how missing data affects SFs of solar wind turbulence. We find that linear interpolation systematically underestimates the true structure function, and we introduce a simple, empirically-derived correction factor to address this bias. Learned from data from a single spacecraft, the correction generalizes well to solar wind measured elsewhere in the heliosphere. Compared to conventional gap handling methods, our approach reduces the mean error for missing data fractions above\textbf{ }20\%, and the overall error is reduced by nearly 50\% when averaged across all missing fractions tested. We apply the correction to Voyager intervals from the inner heliosheath and local interstellar medium (60-85\% missing) and recover spectral indices consistent with previous studies. The correction factor is released as an open-source Python package, enabling more accurate analysis of scaling in gapped solar wind datasets.

\end{abstract}

\section{Introduction}

\subsection{Gaps in solar wind time series}
The solar wind is a supersonic plasma that continuously flows outward from the Sun \citep{parker1958}. As well as playing a key part in our understanding of space weather, the solar wind has a special role in astrophysics more broadly. It is the only system in which we can study \textit{in situ}, using spacecraft, the cosmically ubiquitous phenomenon of weakly collisional plasma turbulence \citep{Bruno2013, matthaeus2021}. Unfortunately, as with countless other datasets in the Earth and space sciences, time series of the solar wind are often plagued by data gaps. Telemetry constraints, instrument failures and data filtering/conditioning all result in time series of plasma and magnetic field measurements that are variably incomplete. For example, data from the two Voyager spacecraft, which provide our only measurements of the outer heliosphere and interstellar medium, have daily gaps of 12-16 hours due to the limited communication with ground stations from such distances \citep{Gallana2016, matsumoto2016voyager}. Our record of the solar wind at Mars is also regularly contaminated by multi-hour gaps, during the periods in which the MAVEN spacecraft passes inside the planet's bow shock \citep{azari2024stats}. These and other significantly fragmented datasets are highlighted in \cref{tab:spacecraft_review}.

\begin{table*}
\begin{tabular}{|l|l|l|l|} \hline 
     Dataset&  Distance from the Sun& Typical \% of magnetic field data missing& Reference\\ \hline\hline 
     Helios 1 \& 2 &  0.3-1 au& 50-60\%  & \cite{venzmer2018}\\ \hline 
     OMNI &  1 au & 67\% until 1995, 8\% thereafter  & \cite{Lockwood2019}\\ \hline 
     MAVEN&  1.5 au (Mars orbit)& 80\%& \cite{henderson2025}\\ \hline
     Voyagers 1 \& 2&  1-140 au& 70\% in the outer heliosphere& \citet{Fraternale2019ApJ}\\ \hline 
\end{tabular}
    \caption{Description of solar wind datasets that are particularly affected by missing data (au = astronomical units). Note that OMNI is a compilation of data from a range of spacecraft at 1 au; the increase in availability after 1995 was due to the Wind spacecraft coming online \textbf{\citep{omniweb}.}}
\label{tab:spacecraft_review}
\end{table*}

The aforementioned causes of gaps in solar wind datasets are such that, generally speaking, we can treat the data in these gaps as ``missing at random''.\footnote{I.e., the presence of gaps is unrelated to the variables of interest\textemdash but is related to external factors, such as the time of day in the case of periodic gaps \citep{little2019statistical}.} While this reduces some potential for bias in the corresponding statistics, large amounts of missing data remain a challenge for various heliophysical analyses. For example, gaps hinder our ability to forecast space weather events \citep{kataoka2021, smith2022}, to understand the coupling of the solar wind with planetary magnetospheres \citep{Magrini2017, Lockwood2019, azari2024stats}, or to study plasma turbulence using scale-dependent statistics \citep{Gallana2016, fraternale2017thesis, dorseth2024}. These statistics are also needed for the validation of turbulence transport models, which are essential for understanding processes such as the propagation of cosmic rays through the heliosphere \citep[e.g.,][]{bieber1996, engelbrecht2013, engelbrecht2022}.

Therefore, it is crucial to investigate whether we can increase the amount of reliable information that can be extracted from such sparse yet scientifically invaluable datasets. In this study we focus on the extraction of robust turbulence statistics\textemdash specifically, the structure function, which we now describe.

\subsection{Structure functions: intuition and advantages}

The statistic most synonymous with turbulence studies is the power spectrum $E(k)$, and in particular, the scaling-law prediction of $E(k)\propto k^{-5/3}$ made by \citet{kolmogorov_41} for the inertial range. However, in that paper, Kolmogorov also introduced the \textit{moments of the increments} $\langle|\Delta x|^p\rangle$ of a turbulent quantity $x$. These statistics, later termed structure functions (SFs) \citep{yaglom1957, obukhov1959}, were also predicted to follow specific scaling-laws for turbulent signals, and since \citet{burlaga_91} have also been used extensively to study fluctuations in the solar wind, especially with regards to the phenomenon of \textit{intermittency} \citep[e.g.][]{TuMarsch1995, BiskampBook, carbone1996, ruzmaikin1995}. Before describing intermittency and the theoretical properties of SFs in this context, we first highlight the three main practical advantages that establish their role alongside power spectra in scaling analyses.

The first advantage offered by the SF is that it allows for direct identification of the range of scales (in time or space) that contribute to the variation of a fluctuating quantity, without needing to translate to or from the frequency domain \citep{yaglom1957, simonetti1985, schulz1981, dewit2013}. The SF clearly visualizes the change in distributional properties (e.g., increasing variance, in the case of the second-order SF) of fluctuations as the separation between measurements increases. The second advantage is that the SF is less sensitive to violations of stationarity \citep{schulz1981, Uribe2021}, compared with the power spectrum and autocorrelation function. This is particularly important for studying the solar wind, where transients such as co-rotating interaction regions and shocks regularly cause non-stationarity of the data \citep{JagarlamudiApJ19}. Thirdly, unlike with the power spectrum, the SF can be straightforwardly calculated from irregularly-sampled time series \citep{cho2001horizontal}.

This final point, while mathematically true, says nothing of the \textit{robustness} of SFs to data gaps\textemdash a property that, as we describe in the next section, has received little attention. In light of this, in this paper we investigate, and demonstrate a method of addressing, the effect of data gaps on the second-order SF. In the remainder of this section, we provide the formal mathematical definition for the SF and its interpretation in a turbulence context, followed by a review of gap-handling methods in the solar wind literature.

\subsection{SFs: definition and turbulence theory}
The $p$th order SF at lag $\tau$ gives the $p$th moment of the distribution of increments at that lag, $\mathcal{P}(\Delta x_\tau)$, where $\Delta x_\tau(t)=x(t+\tau)-x(t)$ is the increment and $x(t)$ is the value of some (scalar or vector) variable $x$ at time $t$ (or equivalently, at a point in space $\mathbf{r}$)

\begin{equation}\label{eq:SF_theory}
\begin{split}
        S_{p}(\tau) & =\int^{\infty}_{-\infty}|\Delta x_\tau|^p\mathcal{P}(\Delta x_\tau) d(\Delta x_\tau), \\ & =\langle|\Delta x_\tau|^p\rangle,
\end{split}
\end{equation}

where angle brackets denote an ensemble average. For an ergodic process, this can be replaced with a time average

\begin{equation}\label{eq:sf_mom}
    S_{p}(\tau)=\frac{1}{N(\tau)}\sum_{i=1}^{N(\tau)}|x(t_i+\tau)-x(t_i)|^p,
\end{equation}

where $N(\tau)$ represents the sample size at a given $\tau$. If either $x(t_i)$ or $x(t_i+\tau)$ is missing due to a data gap, the corresponding increment is excluded, reducing the effective sample size. In addition to assuming ergodicity, the calculation of the SF assumes stationarity of the increments \citep{yaglom2004}. When converting from temporal to spatial lags, it is also assumed that Taylor's hypothesis is valid \citep{taylor1938}. These assumptions\textemdash ergodicity, stationarity, and Taylor's hypothesis\textemdash are commonly made in solar wind studies, though their validity varies depending on the context \citep{M&G1982_stationarity, JagarlamudiApJ19, klein2014violation, Isaacs2015}.

Theoretically, $S_2(\tau)\rightarrow 2\sigma^2$ as $\tau\rightarrow \infty$, where $\sigma^2$ is the variance of $x$. The scale at which $S_2(\tau)$ flattens is approximately equal to the correlation length, a measure of the outer scale of the system.\footnote{Typically, the correlation length is calculated using the autocorrelation, $R(\tau)$. This requires weak stationarity of the signal, and gives the relation $S_2(\tau)/\sigma^2=2[1-R(\tau)]$.} The identification of these so-called ``characteristic scales" is one of the key applications of $S_2(\tau)$ in astrophysics, where it is commonly employed to study the light curves of active galactic nuclei \citep[e.g.,][]{kozlowski2016revisiting, de2022}. In geostatistics, $S_2(\tau)$ is referred to as the variogram and is used as a model of spatial variation in the interpolation scheme known as kriging \citep{matheron1963, webster2007geostatistics}.

The SF has special significance in the analysis of self-similar processes. For a process that exhibits fractal behavior, we observe $S_p\propto\tau^{\zeta(p)}$, where $\zeta(p)$ is a straight line for monofractal behavior and a nonlinear function for multifractal behavior \citep{Frisch1995}.\footnote{The scaling exponent $\zeta$ is related to the Hurst exponent $H$ by $\zeta(p)=pH$.}  In the case of turbulence, classical theory predicts $\zeta(p)={p/3}$ in the inertial range \citep[]{kolmogorov_41, Frisch1995}. Under certain assumptions\footnote{Weak stationarity, zero mean, frequency range from 0 to $\infty$, and $1<\beta<3$ \citep[Appendix B]{emmanoulopoulos2010}.}, we can relate this power-law scaling to that of the power spectrum $E(k)\propto k^{-\beta}$ via $\beta=\zeta(2)+1$ \citep{pope_2000}, giving the famous $\beta=5/3$ power law of turbulence \citep{kolmogorov_41}. This relationship has been exploited in multiple solar wind studies by converting the SF into an ``equivalent spectrum" \citep{Chasapis_2017, Chhiber2018, Roberts2022, thepthong2024scale}, thereby allowing for direct comparison with spectral results.

Therefore, statistical analysis of turbulence often involves fitting power laws to SFs and comparing the corresponding scaling exponents with theoretical predictions\footnote{As has been noted elsewhere \citep{emmanoulopoulos2010}, this common analysis is statistically problematic, due to violations of linear regression assumptions of independent and Gaussian-distributed errors, which lead to underestimates of the error associated with the fits. This issue is not critical to the present work, but does warrant consideration in future practical implementations of our correction.} \citep{BRUNO2007}. This includes their wide application to a range of astrophysical flows, including the interstellar medium \citep[e.g.,][]{Boldyrev_2002, Padoan_2003}, intracluster medium \citep[e.g.,][]{Li_2020, gatuzz2023}, and, as highlighted in the present work, the solar wind \citep[e.g.,][]{Horbury1997, Bigazzi2006, Chen_2012, pei2016influence}. As well as the aforementioned practical advantages, higher-order SFs are particularly well-suited to probing increment distributions in finer detail.\footnote{This said, higher-order spectra have been developed and are reported to be more adept than SFs at handling the effects of non-stationarity and large-scale structures \citep{huang2011, carbone2018}.} In particular, the \textit{intermittency} of turbulent fluctuations is of interest. Intermittency refers to a greater propensity for particularly large fluctuations, i.e., a heavy-tailed probability distribution. This can be directly quantified via the kurtosis of the distribution, which is given by the normalized fourth-order SF $S_4(\tau)/S_2(\tau)$ \citep{Frisch1995}. Although intermittency is a well-known phenomenon in turbulent flows, it is of particular interest in the solar wind, due to its role in understanding the sites and mechanisms of energy dissipation in weakly collisional plasmas \citep{tenbarge2013current, MattEA15-philtran,Chhiber2018,bruno2019}.  Therefore, SFs are a uniquely valuable tool for probing the physics of space plasmas \citep{dewit2013, Bruno2013}. In order to maximize the amount of reliable science that can be extracted from them, it is vital that we better understand their much-touted robustness for irregularly-sampled time series.

\subsection{Current approaches to handling data gaps}

The power spectrum has received much more attention than the SF with regards to the effect of gaps, due to the more immediate obstacle gaps pose to its calculation, and the ubiquity of the power spectrum across science and engineering. We briefly review this work now. Spectral analyses of the solar wind typically handle small gaps in time series (around a few percent in length) using linear interpolation \citep[e.g.,][]{vrvsnak2007coronal, ChenApJS20, carbone2021}. When faced with larger gaps, one could simply work with gap-free subsets. However, this is often inadequate when seeking to analyze a particularly wide range of scales in order to obtain a complete picture of the turbulent energy cascade \citep{dorseth2024}. Therefore, the performance of various interpolation methods and alternative spectral estimators, including the Blackman-Tukey method \citep{blackman1958} and Lomb-Scargle periodogram\textbf{ \citep{lomb1976least, Lee_2020}} have been compared \citep{munteanu_2016, Magrini2017}. (For studies outside of heliophysics, see \citet{dey2021effect, mao2024novel, arevalo2012, Babu2010}.) Using such techniques, accurate spectral estimation has been claimed for solar wind datasets with missing fractions of up to 50\% \citep{dorseth2024}, 70\% \citep{Gallana2016, fraternale2017thesis, Fraternale2019ApJ} or even 80\% \citep{mckee2020analysis}.

Unlike the power spectrum, by remaining in the time domain, the SF is immediately calculable from irregularly-sampled time series (see \cref{eq:sf_mom}). This has led to confident use of the statistic when faced with small gaps (see, e.g., \citet{Horbury1997, Bigazzi2006}, for solar wind studies, and \citet{takahashi2000, zhang2002, seta2023rotation, mckinven2023} for astrophysical studies). More recently, however, SFs have been claimed to be robust for much larger missing fractions. For example, in atmospheric physics, a variant of the SF based on the Haar wavelet was recommended as a spectral estimator for sparse turbulent-like signals \citep{mossad2024}. For solar wind data, a pair of studies have suggested that up to 70\% data loss can have only a negligible effect on statistics derived from the SF, which we discuss presently. \citet{Burger2023} created a synthetic dataset with known spectral properties, and then decimated it according to Voyager gap distributions of 64\% and 68\% sparsity. The resultant spectral index and power in the inertial range, as well as the correlation time, deviated by no more than 5\% from their true values, as shown in their Table 1. A similar result was also found for data gaps from the IMP and ACE spacecraft \citep[Table 1]{burger2022}. However, these results were only for a single synthetic dataset, and, as acknowledged by the authors, one would expect ``somewhat'' different results for different realizations of their turbulence simulation.

\citet{Fraternale2019ApJ}, on the other hand, reported that both the amount and distribution of missing data in Voyager datasets makes computation of the SF ``nontrivial". They showed that the periodic gaps present in these time series lead to regular oscillations in the sample size, which in turn produce artifacts in time-domain statistics such as the SF.\footnote{This behavior was illustrated for the closely-related autocorrelation function by \citet[][Supplementary Information]{Gallana2016} and \citet{dorseth2024}.} Proceeding to calculate the SF without any interpolation, the authors took this behavior into account by using a statistical significance threshold based on relative sample size when calculating the spectral index. The statistical convergence of SFs affected by gaps was explored in a later work \citep[][Appendix B]{fraternale2021}.

In an astrophysical context, \citet{emmanoulopoulos2010} also challenged the assumption that SFs are immune to missing data, as part of a wide-ranging critique of over-interpreting SFs when studying the variability of blazars. In a brief qualitative analysis, it was shown that gaps severely affect SF estimates in an unpredictable manner that is dependent on the specific time series (see their Figure 12). Therefore, they concluded that extensive simulation is necessary to account for this behavior.

In order to inform strategies for more robust SF estimation, we perform such simulation and thereby thoroughly test the resilience of SFs to gaps. We empirically derive the average bias, with and without linear interpolation\textbf{,} and investigate whether a simple correction can be made to ``de-bias'' these errors. As our ground truth, we use magnetic field measurements from the Parker Solar Probe and Wind spacecraft, as described in Section \ref{sec:data}. In Section \ref{sec:method}, we detail the extensive gap simulation of these intervals, followed in Section \ref{sec:results} by the results on a set of case studies and the overall statistical picture, as well as an application to Voyager intervals. Section \ref{sec:discuss} interprets the results in the context of other approaches to gap-handling, including future work and practical considerations of using the correction. Section \ref{sec:concl} summarizes the study before describing how to access the correction.

\section{Data}\label{sec:data}

\begin{table*}
    \begin{tabular}{|l|l|l|l|l|} \hline 
          Dataset&Spacecraft&  Intervals $\times$ versions&Mean $\delta b$ (nT)&Mean $\lambda_C$ (km)\\ \hline \hline 
          Training&Parker Solar Probe&  10,731 $\times$ 25& 20&$10^5$\\ \hline 
          Test&Wind& 589 $\times$ 25& 4&$10^6$\\ \hline 
          Application& Voyager 1& 2 $\times$ 1& 0.02&$10^8$\\ \hline
    \end{tabular}
    \caption{Description of each of the datasets used in this study. In each case, the intervals consist of the full vector (3-component) magnetic field data. ``Versions'' refers to the number of duplicates that are created from the original complete intervals, which are then artificially gapped in different ways (the Application intervals originally contain gaps, hence these are not duplicated). Average values of the turbulence parameters of rms fluctuation magnitude $\delta b$ and correlation length $\lambda_C$ are obtained from the datasets under study for PSP and Wind, and from \citet{fraternale2021} for Voyager 1.}
    \label{tab:metadata}
\end{table*}

The aforementioned investigations into the effects of gaps on spectral estimation used simulated time series, real solar wind intervals, or a combination of the two. Here, we restrict ourselves to real intervals, so as to avoid the simplifying assumptions when working with simulated data, such as Gaussianity. We also use considerably more intervals than the other works cited. 

The specific datasets used in this work, each from different spacecraft, are summarized in \cref{tab:metadata}. They comprise a training set to calculate the biases, a test set to verify the biases generalize, and an application set to demonstrate the practical use of our de-biasing algorithm. For each dataset we extract the full vector magnetic field time series collected by the spacecraft's magnetometer instrument and divide it into intervals according to the standardization procedure described at the end of this section.

For the training set, we use data from Parker Solar Probe (PSP), a mission launched in 2018 to study the origins of the solar wind by flying very close to the Sun (less than 10 solar radii at closest approach) \citep{Fox2016}. This data provides us with the long, continuous time series required to perform comprehensive gap simulation for a range of turbulence realizations. We use data from the years 2019-2023 as measured by its fluxgate magnetometer instrument at a native cadence of 256 samples/second \citep{Bale2016, psp_data}. Following standardization (discussed below), this yields 10,731 intervals for training the de-biasing algorithm.

For the test set, we use data from Wind. Wind is a spacecraft situated at the L1 Lagrange point that has been continuously measuring the near-Earth solar wind since May 2004, and has contributed significantly to our understanding of turbulence in the solar wind \citep[e.g.,][]{Woodham2018,Verdini_2018,Wilson2020}. We use data collected by the Magnetic Field Experiment \citep{Lepping1995, wind_data} at a native cadence of 11 samples/second during the period March-December 2016. Following standardization, this yields 589 intervals for evaluating the algorithm.

Finally, we apply our de-biasing algorithm to one of the gap-afflicted datasets identified in \cref{tab:spacecraft_review} as an example of how it could be used in practice; this is our `application' data.  We take two highly fragmented, approximately week-long intervals from the Magnetic Field Experiment onboard Voyager 1 \citep{voyager_magnetometer,voyager_data}: one from 118 au in the inner heliosheath with 65\% missing data, and one from 154 au in the local interstellar medium with 86\% missing data. Both are measured at native cadence of 1 sample every 48 seconds (0.021 samples/second). Voyager data from the outer heliosphere and interstellar medium represents our only in situ measurements of the solar wind in these regions. Given its high percentage of missing data, this makes the Voyager datasets a clear example of where improved statistical estimation could reap significant scientific benefits \citep{Gallana2016, Fraternale2019ApJ, BurlagaApJ2018}. 

However, it would be naive to simply learn the SF biases from gapped PSP data, and then directly apply it to data from the edge of the heliosphere\textemdash a very different physical regime.\textbf{  }For example, the correlation time \citep{Cuesta2022} and the magnetic fluctuation amplitude \citep{ChenApJS20} show a large degree of variability at different radial distances from the Sun\textemdash and even within a localized region of space or a single orbit \citep{maruca2023trans}.\textbf{ }The typical values of these two quantities for each of our spacecraft are given in \cref{tab:metadata}, showing that, for example, correlation times vary by three orders of magnitude between the inner and outer heliosphere. Therefore, we apply a dual standardization to each interval, in order to improve the likelihood that our results generalize across various turbulent regimes.

Firstly, to account for the local correlation time\textemdash and therefore any scale-dependence of bias in the SF\textemdash we standardize each interval to a consistent number of correlation times over a consistent number of points. In this work, each interval is made to contain 10 correlation times across 10,000 points. Secondly, we standardize power levels (i.e., fluctuation amplitudes) by normalizing each interval to have a mean of 0 and variance of 1. This makes the size of the biases introduced by gaps more consistent. Therefore, we have normalized both ``axes'' of each interval, with the aim of making our analysis system-agnostic. More detail of this process and an illustrated example are given in the Appendix.

\section{Method} \label{sec:method}

\begin{figure*}
  \centering
  \includegraphics[width=0.8\textwidth]{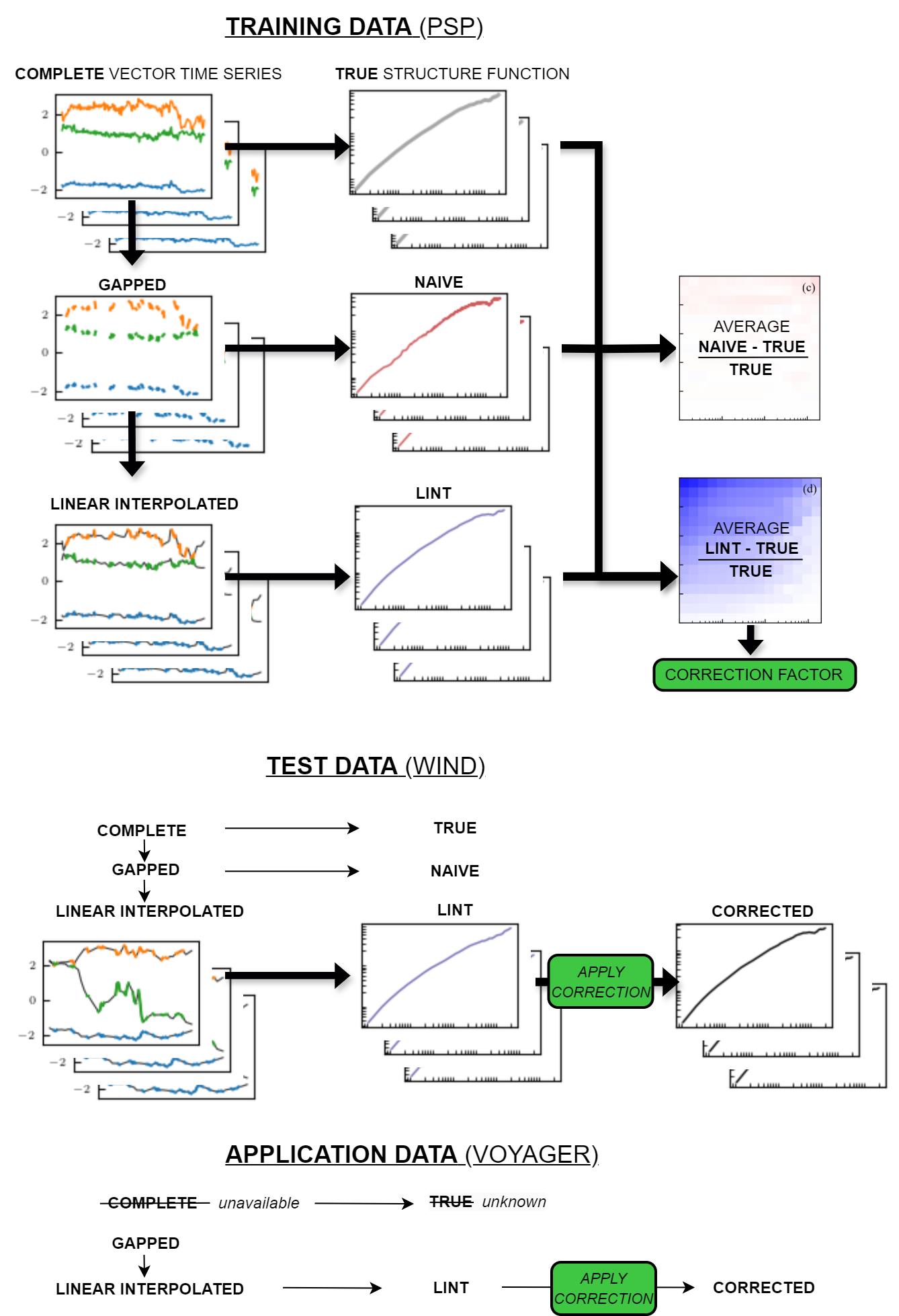}
  \caption{Flowchart demonstrating the method of the study. Prior to computing the SFs, the complete interval of each spacecraft, or gapped interval in the case of Voyager when the complete data is not available, is first standardized according the procedure outlined in the Appendix.}
  \label{fig:flowchart}
\end{figure*}
%\clearpage

\textit{This section is summarized in the flowchart in \cref{fig:flowchart}.}

We compute the vector second-order SF (hereafter simply SF) for each complete, standardized 3$\times$10,000 magnetic field time series using  \cref{eq:sf_mom} with $p=2$. This is our ground-truth, $S_2(\tau)$. We compute each SF up to a lag of 20\% of each standardized interval\textemdash containing 10 correlation lengths\textemdash resulting in the full SF spanning two correlation lengths for every interval in all three datasets. This standardized range of scales also ensures a reasonable (pre-gapping) sample size at each lag.

We then create 25 copies of each time series and introduce random gaps. As well as increasing the sample size for our statistical analysis, duplicating the intervals allows us to study the effect of different gap distributions on the same interval, as will be demonstrated in \cref{fig:case_study_gapping}. We also simulate gap distributions that combine two types of missing data: uniformly distributed (individual) missing points, and contiguous ``chunks'' of missing points. The differing effects of the two types have been demonstrated by \citet{emmanoulopoulos2010} and \citet{dorseth2024}. In both of these studies, uniformly distributed gaps were shown to merely add noise to the SF\textemdash an effect that is relatively easily ameliorated by linear interpolation. Contiguous gaps, on the other hand, such as the multi-hour gaps in the Voyager and MAVEN datasets, significantly distort the shape of the SF. For greater fidelity to spacecraft data, we apply both gap types to the same time series, but emphasize the contiguous case. Specifically, for a given total gap percentage (TGP), which we vary up to as much as 95\%, we require that at least 70\% of that amount must be removed via contiguous chunks. The exact proportions are chosen randomly. 
%Another key difference from previous gap studies is that we decimate individual intervals multiple times using different random gap configurations.

From these gapped intervals, we re-compute our SF estimates, denoted using hat notation as $\hat{S}_2(\tau)$. Firstly, we use \cref{eq:sf_mom} without any interpolation: this is our ``naive'' estimate, \naive, following the common approach in the literature \citep[e.g.,][]{Horbury1997, Bigazzi2006, mckinven2023}. Secondly, we apply linear interpolation, calling this our ``LINT'' estimate, \lint. This allows us to study the behavior of this very simple and common imputation method.
%, albeit one that isn't strictly necessary for time-domain statistics.

The existing literature on quantifying the effect of gaps on SFs has been limited to the effect on derived statistics, i.e., the inertial range slope and correlation length \citep{Burger2023}. However, we note the importance of accurately estimating the amplitude and shape of the entire SF curve. For example, the entire SF is required to compute the kurtosis. Therefore, we evaluate the error of the overall SF. We quantify this error for a given $\tau$ using the percentage error, PE, defined as follows:

\begin{equation}\label{eq:PE}
    \text{PE}(\tau) =
            \frac{\hat{S}_2(\tau) -S_2(\tau)}
           {S_2(\tau)}\times100.
\end{equation}

The overall error of an SF estimate is given by the mean absolute percentage error across all lags, MAPE,

\begin{equation}\label{eq:MAPE}
    \text{MAPE} = \frac{1}{n_{\tau}} \sum_{\tau=1}^{n_{\tau}}|\text{PE}(\tau)|,
\end{equation}

where $n_{\tau}$ is the number of lags over which the SF has been computed. Later in this paper, when creating our correction factor, we will define additional error metrics.

\section{Results} \label{sec:results}
To begin, it is useful to recall the basic issue at hand: gaps in a time series reduce the sample size $N(\tau)$ of each lag distribution $\mathcal{P}(\Delta x_\tau)$. Unlike simply having a shorter interval, gaps result in different lags being depleted to different degrees, depending on the size and location of the gaps. Uniformly distributed gaps will tend to reduce $N(\tau)$ uniformly across all lags; whereas contiguous gaps, which we emphasize here, result in \textit{uneven} depletion of the distributions. In either case, reduction in sample size affects the variance of said distribution, and therefore the value of the SF. %The exact way in which each variance is affected is ambiguous: lags could be removed from the edges of the distributions, or the middle. 
Here we show examples of this effect on an individual time series, before proceeding to the overall trends.

\subsection{Effect of gaps: case studies}

\begin{figure*}
    \centering
    \includegraphics[width=7in]{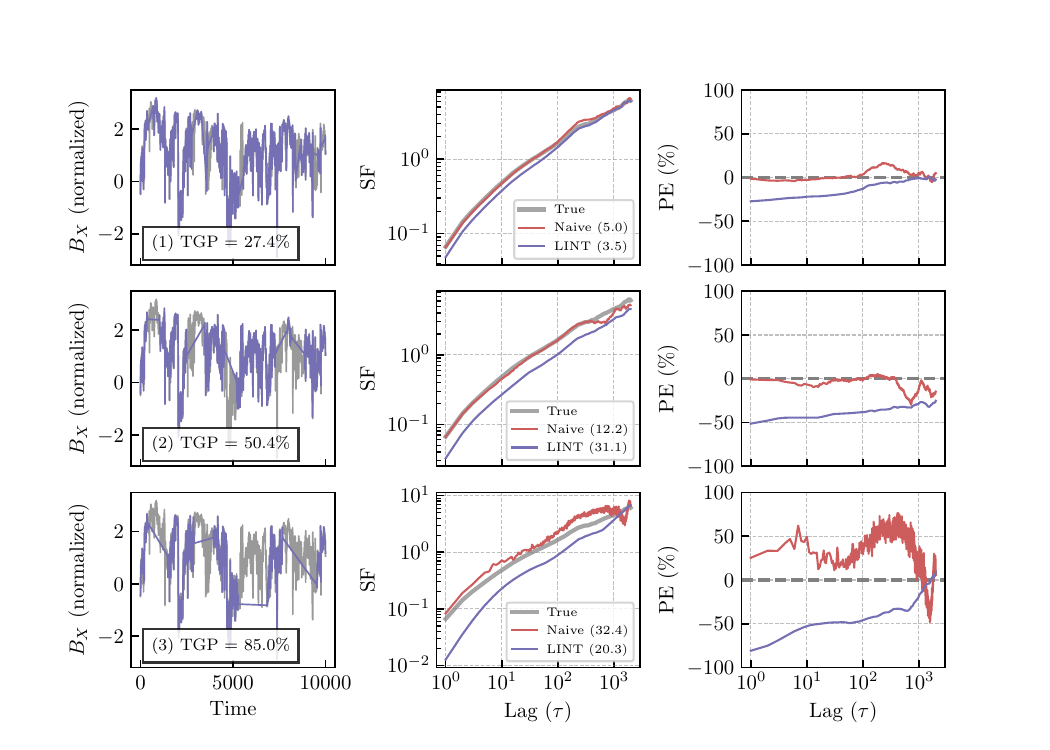}
    \caption{\textbf{Examples} of the effect of increasing amounts of missing data on the SF of \textbf{three gapped versions of a single Wind interval}. The left-hand column shows the original complete interval (grey), and the interpolated gapped interval (purple). Only one of the three vector components used in the calculation is shown, for visualization purposes. The middle column shows the SF from the complete interval (``true'', thick grey), as well as the two estimates: \naive\ (``naive'', red), and \lint\ (``LINT'', purple). The mean absolute percentage error of each estimate is given in brackets. The percentage errors as a function of lag are given in the right-hand column.}
    \label{fig:case_study_gapping}
\end{figure*}

\cref{fig:case_study_gapping} shows a case study of the effect of gaps on an interval from the Wind spacecraft, gapped in three different ways. (We discuss case studies from our Wind test dataset, rather than our PSP training set, in order to later compare with the corresponding corrected versions.)

For the Naive approach, whereby we simply ignore the gaps when estimating $S_2(\tau)$, we observe minimal distortion in this SF for both (1) 27.4\% and (2) 50.4\% missing. Looking at the percentage error as a function of lag, the error is close to 0 at least up to lag 100, beyond which we see over- or under-estimation of about 10-20\% at these larger scales. At 85\% missing data (3), we see more significant distortion across the entire SF, with the introduction of both both low-frequency oscillations and high-frequency noise. These distortions lead to both under- and over-estimation of the true curve.

In contrast, the LINT approach shows a consistent under-estimation of the SF that increases with increasing TGP (the reasons for which we discuss in the next section). While the magnitude of this error is larger than that of \naive for interval versions (1) and (2), it is also much more consistent. Particularly for 50.4\% missing data, the PE is relatively flat across all lags: we can see that \lint\ retains the true shape of $S_2(\tau)$ extremely well, despite earning the higher overall MAPE of the two methods. Interval version (3) has strong variation in PE across lags, but unlike \naive, the trend is reasonably monotonic. This systematic trend suggests that we may be able to learn the typical bias as a function of lag and percentage missing, and thereby `push' the LINT estimate back towards its true values.

(We note that the errors observed in \cref{fig:case_study_gapping} are not as extreme as those illustrated in Fig. 12 of \citet{emmanoulopoulos2010}, despite similar missing fractions. This is likely because their intervals were 1/5 the length of ours and therefore more affected by gaps.) 
%However, that study shows the trend for high and low frequency fluctuations to be introduced by uniform and contiguous gaps, respectively. 

We now move to a statistical analysis to look at overall trends in behavior across the entire PSP training set, which contains many different realizations of solar wind turbulence and gap distributions.

\subsection{Effect of gaps: statistical analysis}

\begin{figure*}[!h]
    \centering
    \includegraphics[width=5.5in]{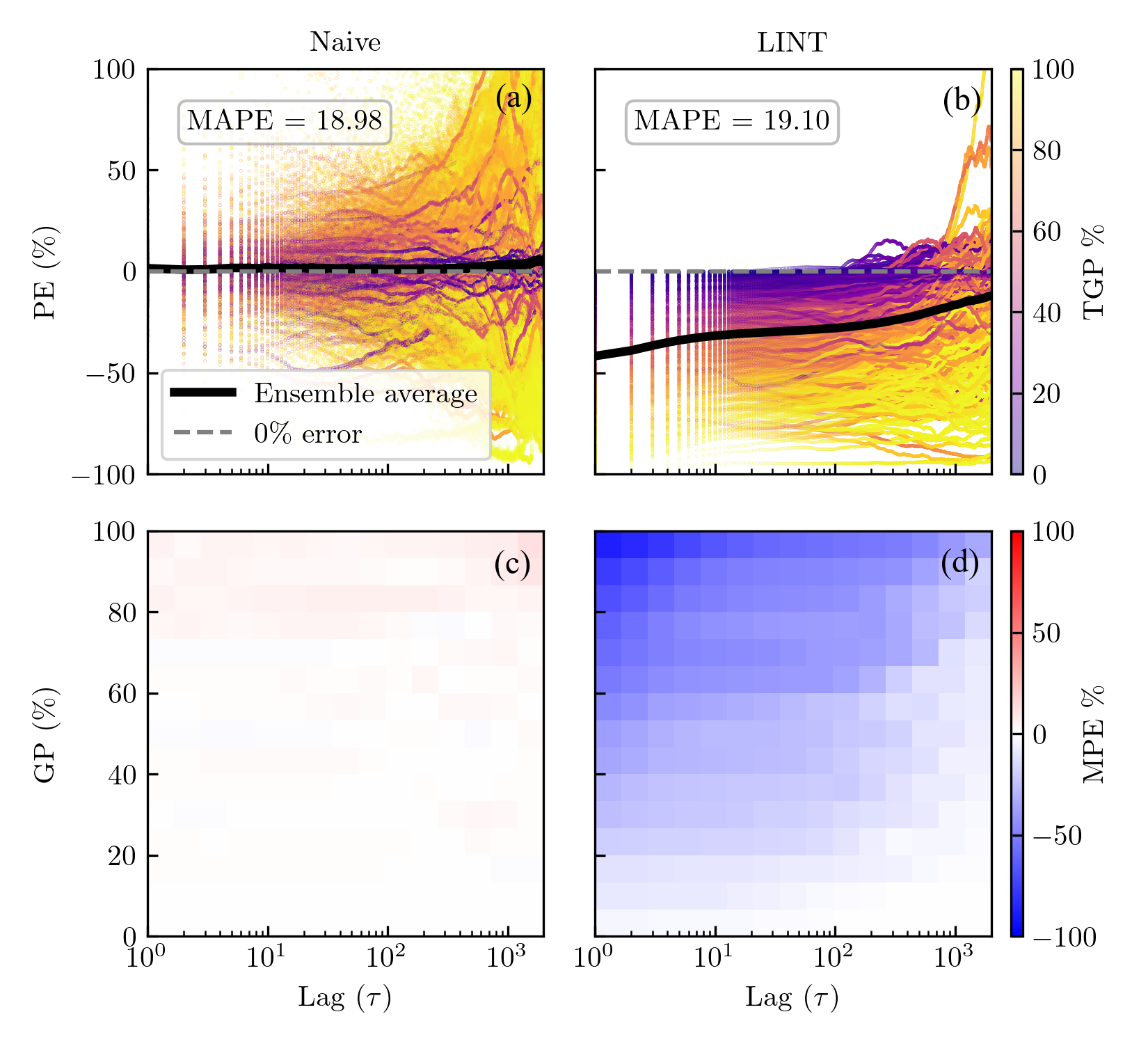}
    \caption{Two representations of relative error as a function of lag and missing fraction, as calculated from the PSP training set. Results for the \naive\ are given in (a) and (c) and \lint\ in (b) and (d). Percentage error (PE) trendlines are given in (a) and (b) for a subset of 775 intervals, as in the third column of \cref{fig:case_study_gapping}, colored according to the total gap percentage (TGP) of that interval. The black lines show an ensemble average of each trendline (see \cref{eq:MPE}). A 0\% error line (dashed grey) is shown for reference. Subplots (c) and (d) show mean percentage error (MPE) for each combination of lag-specific gap percentage (GP, 15 linear bins) and lag (15 logarithmic bins)\textemdash see \cref{eq:mpe_bin} for the calculation. These heatmaps are calculated from the full training set of $\approx$250,000 intervals.}
    \label{fig:train_psp_error_trend}
\end{figure*}

The statistical analysis visualized in \cref{fig:train_psp_error_trend} shows the SF estimation errors from our large PSP training set. In addition to the PE for each individual SF estimate, we also show in the top row the mean percent error, or ensemble average, across all SFs at each lag $\tau$, where $N$ is the total number of SFs estimated:

\begin{equation}\label{eq:MPE}
    \text{MPE}(\tau) = \frac{1}{N} \sum_{i=1}^{N}\text{PE}(\tau)_i.
\end{equation}

Referring to this ensemble average, we see in \cref{fig:train_psp_error_trend}a that the average error remains close to zero for all lags. This clearly demonstrates the unbiased nature of the naive SF estimator, such that its expected value $E[\hat{S}_2^\text{naive}(\tau)]=S_2(\tau)$. Despite the often very large errors caused by gaps at large lags, often approaching 100\% underestimation and exceeding 100\% overestimation, there is no consistent bias away from the true value in either direction. Moreover, while greater errors tend to be correlated to TGP (note the error vs. color of the trendlines), there is unpredictable variability in this dependence, as well as in the amplitude of the error. The fact that high TGP can result in only low or moderate errors in \naive\ was shown in the case study of 50\% missing in \cref{fig:case_study_gapping}, as well as the results of \citet{Burger2023} where TGPs of 64\% and 68\% caused negligible changes in SF-derived statistics.

\cref{fig:train_psp_error_trend}b shows the errors from linearly interpolating the gaps. We see a very similar overall absolute error for the set of intervals (19.1 vs. 18.98), but a very different picture with regards to the direction of this error. Up to $\tau=100$ (10\% of a correlation time), this estimator shows consistent underestimation\textemdash$E[\hat{S}_2^\text{lint}(\tau)]< S_2(\tau)$\textemdash that increases reasonably smoothly with TGP. This decrease in the variance of the lag distributions $\mathcal{P}(\Delta x_\tau)$ is expected given that drawing straight lines across gaps is equivalent to smoothing the time series and removing variation; the same phenomenon has been observed in the power spectrum \citep{Fraternale2019ApJ}. We see greater underestimation at small lags because, in spite of having a larger sample size, these distributions are much more distorted by long periods of gaps. When $\tau$ is small, it is more likely that both of the values $x(t), x(t+\tau)$ in the difference $x(t)-x(t+\tau)$ occur on the same interpolated line, and the longer this line, the smaller this new difference will be. This results in a dramatic shift of the increment distribution $\mathcal{P}(\Delta x_\tau)$ towards the center and therefore an excessive decrease in the variance of this distribution, corresponding to a reduction in the value of the SF at that scale.\footnote{We also examined the error of two alternative estimators of the SF, Cressie-Hawkins and Dowd \citep{cressie1980,dowd1984variogram,webster2007geostatistics}. These estimators, from geostatistics, are designed to be more robust to outliers and skew than the traditional estimator we use here (also known as the Matheron method-of-moments estimator). It was found that both of these estimators were in fact more sensitive to gaps, producing larger errors, and did not have the same unbiased property of the traditional estimator.}

Above $\tau\approx100$, the ensemble average error of \lint\ gets closer to zero as we begin to see overestimation for some intervals at these larger lags, but it remains negative. At this point, the correlation between TGP and error also becomes weaker.

\cref{eq:MPE} is a 1D average error. We also calculate a 2D error, as a function of lag and the gap percentage \textit{at a given lag}, $\text{GP}(\tau)$\textemdash as previously noted, this is dependent on both the size and location of gaps. We bin these two variables and calculate the mean PE of the \lint\ estimate in each bin $\mathcal{B}$:

\begin{equation}\label{eq:mpe_bin}
\begin{split}
\text{MPE}(\mathcal{B}) &= \frac{1}{N(\mathcal{B})} \sum_{\tau,\text{GP} \, \in \, \mathcal{B}} \text{PE}_\text{LINT}(\tau,\text{GP}), \\
\end{split}
\end{equation}

where $\mathcal{B}$ is a 2D bin defined according to a range of $\tau$ and GP. Each bin is then colored according to its corresponding $\text{MPE}(\mathcal{B})$, resulting in the heatmaps shown in \cref{fig:train_psp_error_trend}c and d.

\cref{fig:train_psp_error_trend}c shows the binned error of the \naive\ estimates. The almost blank heatmap reinforces the unbiasedness of simply ignoring the gaps. We only see a very small average positive error in the vicinity of $\tau>800$ and GP $>$ 80\% in the top right. For the \lint\ errors in \cref{fig:train_psp_error_trend}d, we see the minimum (near-0) average error at large lags and low missing \% in the bottom right, which then becomes an increasingly negative error (underestimating the SF) at smaller lag and higher missing \%, moving towards the top left of the figure. As already described, smaller lags see stronger underestimation as it is more likely at those scales that most of the increments are computed from interpolated segments.

\subsection{Computation of correction factor}

\begin{figure}
    \centering
    \includegraphics[width=3.5in]{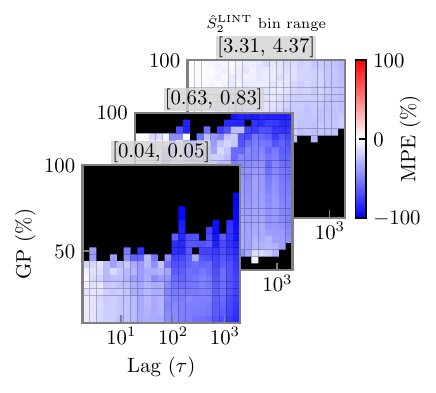}
    \caption{Three slices of the 3D ``error cube'' used for calculation of the final correction factor $\alpha$ using \cref{eq:correction_factor}. The 2D error heatmaps of the \lint\ estimator, as given in \cref{fig:train_psp_error_trend}d, are now additionally computed across 25 bins of the estimated SF value by said estimator. Bin ranges are given in square brackets; the full correction factor uses 25 such bins across all 3 dimensions of GP, lag, and \lint. Black regions indicate unsampled bins. The shifting error as one moves up the SF (across bins) is clear.}
    \label{fig:3d_heatmap}
\end{figure}

% \begin{figure}
%     \centering
%     \includegraphics[width=3.5in]{3d_scatter.pdf}
%     \caption{Illustration of the final correction factor calculation; essentially an extension of the heatmap in \cref{fig:train_psp_error_trend}. Each combination of lag, gap percentage, and power (i.e., the value of the SF at that lag) is binned, and the mean percentage error (MPE) of that bin is computed. These values are then used to derive the final correction factor according to \cref{eq:correction_factor}. The number of points and bins shown here are reduced for illustrative purposes. 25 bins along each dimension are used in the final analysis.}
%     \label{fig:3d_cube}
% \end{figure}

Noting the consistent bias observed in \lint, we investigate whether this bias is in fact consistent enough that it can be utilized to correct any given estimate of the SF from a sparse time series. In other words, we test the ability of a correction factor to ``de-bias'' \lint. Currently, as shown in \cref{fig:train_psp_error_trend}d, we calculate the average error introduced at a given lag ($\tau$) by a given reduction in the sample size at that lag (GP). In order to improve the specificity of this correction, we also calculate error as a function of the value, or ``power", of the estimated SF itself, \lint. We do this by introducing \lint\ as a third variable into \cref{eq:mpe_bin}, resulting in a $25\times25\times25$ `error cube'. The results for a subset of bins along this additional dimension can be seen in \cref{fig:3d_heatmap}. It is clear looking across bins that the average error changes across all three dimensions: as the value of \lint\ increases, the MPE for a given ($\tau$, GP) gets closer to zero. Therefore, adding this additional information should make the corrections more precise (at the expense of smaller sample sizes).

We then convert the MPE in each bin into a multiplicative correction factor, using the following equation:

\begin{equation}\label{eq:correction_factor}
    \alpha(\mathcal{B})=\frac{100}{100+\text{MPE}(\mathcal{B})}.
\end{equation}

Any values of \lint\ in the test set which fall into this bin are then multiplied by $\alpha$ to attempt to return it to its ``true'' value. This gives us our corrected SF:

\begin{equation}\label{eq:corrected_sf}
\hat{S}_2^\text{corr}(\tau, \text{GP})=\alpha(\mathcal{B}) \hat{S}_2^\text{lint}(\tau, \text{GP}).
\end{equation}

For example, a bin with MPE=-30\% (i.e., underestimating the true SF by 30\% on average) will have a corresponding $\alpha\approx1.43$. As a final step, the set of corrections for a given SF are smoothed using a cubic interpolation in log-space. This removes the discontinuities that occur in the corrected SF due to somewhat different values of $\alpha$ for neighboring bins.

\subsection{Validation of correction factor on Wind data}

The Wind test set, once standardized and and gapped 25 different ways, consists of \textbf{14,725} intervals. The key ``hyperparameter'' to tune in developing this correction factor was the number of bins, which represents a trade-off between sample size and specificity when it comes to the precise correction factor for each combination of variables. We trialed 10, 15, 20, and 25 bins. In a similar trade-off, we compared only binning on GP and $\tau$, versus binning along a 3rd dimension of power. The model was evaluated on the test set using the mean MAPE of $\hat{S}_2(\tau)$ across all the intervals in the test set.  

% \begin{table}
%     \centering
%     \begin{tabular}{|l|l|l|} \hline
%           Method & SF MAPE (SD) & Slope APE (SD) \\\hline \hline
%            Naive&  16.8 (16.8)& 12.8 (14.1)\\ \hline
%           LINT & 17.3 (17.8)& 13.0 (13.9)\\ \hline
%   Corrected& 8.7 (8.2)& 9.0 (10.4)\\ \hline
%     \end{tabular}
%     \caption{}
%     \label{tab:test_set_results}
% \end{table}

\begin{figure}
    \centering
    \includegraphics[width=3.5in]{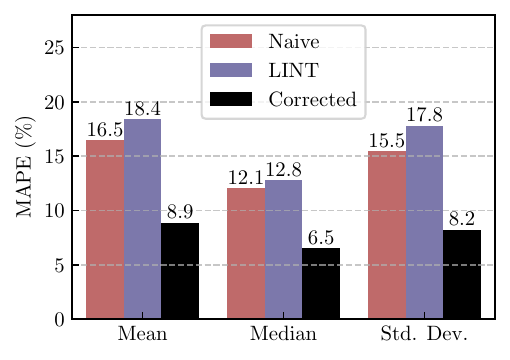}
    \caption{Performance of each method on the Wind test set. SF MAPE is the overall SF estimation error, quantified by the mean MAPE over all SFs. Slope APE is the overall error in the estimated slope of the SF, quantified using the absolute percentage error (APE), averaged over all SFs. SD = standard deviation.}
    \label{fig:test_set_results}
\end{figure}

Using these metrics, it was found that using all 3 variables and 25 bins to compute $\alpha$ gave the best performance. The overall results for this estimator are given in \cref{fig:test_set_results}. We can see that \corr\ indeed improves the overall SF estimation, according to both the mean and median error, with the typical percentage error about half that of the LINT method, and 54\% that of Naive. A similar trend is shown for the standard deviation of error, showing that the correction not only reduces the error, but also reduces the variation in estimates.

\begin{figure*}
    \centering
    \includegraphics[width=7in]{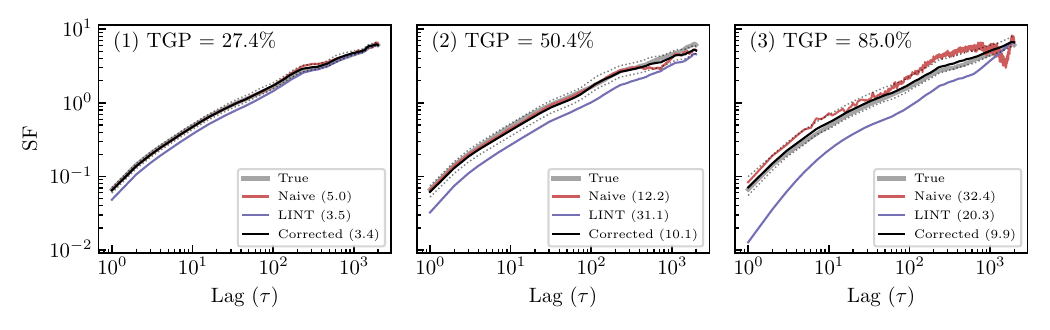}
    \caption{Examples of applying the correction factor to \lint\ to three gapped versions of a single Wind interval (the same interval studied in \cref{fig:case_study_gapping}).  The dotted black lines indicate \corr\ $\pm$ two standard deviations.}
    \label{fig:case_study_correcting}
\end{figure*}

However, it is more useful to understand the performance of each estimator as a function of data sparsity. As before, we start by showing case studies. In \cref{fig:case_study_correcting}, we perform the correction on the same gapped Wind interval studied in \cref{fig:case_study_gapping}. We also provide a measure of uncertainty in the corrections, using the variation in errors used to calculate $\alpha$. Specifically, we obtain the upper and lower limits of \corr\ by adding and subtracting two times the standard deviation of the PEs in each bin $\mathcal{B}$ from $MPE(\mathcal{B})$ in \cref{eq:correction_factor}.

% For interval (1), \corr\ is an improvement on \naive\ but is worse than \lint, having a higher MAPE of 7.2 vs. 2.3. However, as we significantly increase the TGP, \corr\ is superior to both of the original estimators. In interval (2), we can see the correction has achieved precisely its goal: taking advantage of the consistent underestimation but relatively accurate shape maintained by \lint, the correction has essentially just translated the SF upwards to very closely match its true power level. While it has slightly over-corrected at small lags and under-corrected at large lags for this example, the confidence region overlaps with the true SF nicely. Finally, for interval (3) we again see good performance, accurately translating smaller lags upwards by a larger amount than large lags. It still maintains an edge over \naive\ here, even though that estimate is unusually accurate for such a high TGP.

What we observe is that the correction has indeed largely achieved its goal of pushing the LINT curve upwards towards the true SF. At a TGP of 27.4\%, this is only a very marginal improvement on the original LINT estimate (MAPE of 3.4 vs. 3.5). However, as the TGP increases, it becomes a more significant relative gain. In particular, we highlight interval version (3), where the correction has correctly shifted the values at the smallest lags up by a greater degree than at larger lags. This results in an SF that not just has accurate power, but is also much less noisy than the corresponding Naive estimate.

\begin{figure*}
    \centering
    \includegraphics[width=7in]{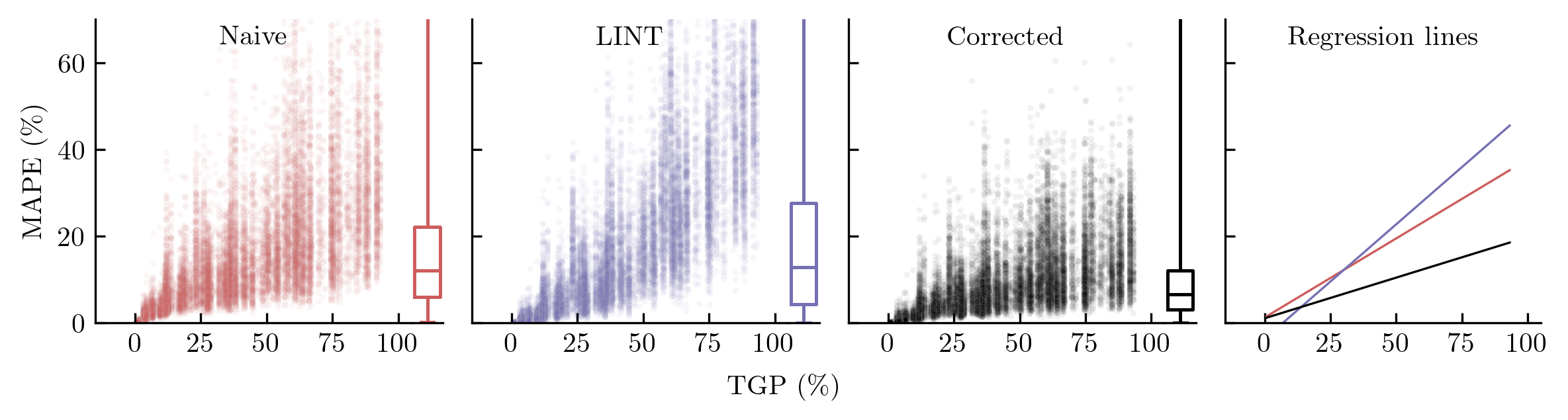}
    \caption{Error as a function of TGP for each of the three SF estimators for the Wind test set. On the right axis of each scatterplot is a boxplot showing the univariate distribution of errors: note some of the points are outside the plotted area in order to show the bulk of the distributions. The final panel shows linear regression lines fitted to each scatterplot. A 99\% confidence region plotted but is smaller than the lines\textemdash note caveats in the text.}
    \label{fig:test_wind_scatterplots_25_bins}
\end{figure*}

The test set errors for each method as a function of missing fraction is given in \cref{fig:test_wind_scatterplots_25_bins}. As expected, all methods show generally increasing MAPE, and increasing variance in MAPE, with greater TGP. While noting that this heteroskedasticity violates one of the assumptions of least squares regression, we fit linear regression lines to give a rough indication of the range of TGP each estimator is best suited for. These fits suggest that \lint\ tends to produce the smallest errors at very low missing fractions ($<10\%$), as well as the largest errors at moderate-to-high TGP ($>40\%$). (This was demonstrated in the case study of interval (2) in \cref{fig:case_study_correcting}: because \lint\ shows more predictable underestimation, it can produce worse estimates than \naive\ for large TGPs.) For TGP$>$20\%, \corr\ tends to show the lowest average error. The gap between methods widens with higher TGP: the scatterplots show that all \corr\ errors remaining below about 50\% all the way up to TGP=95\%. Meanwhile, gaps can lead to errors in \naive\ and \lint\ errors over 60\% as early as $\approx$60\% missing. 

\subsection{Application to Voyager data}

Given the success of our empirical, PSP-derived correction factor, we now apply it to the SFs of two highly sparse intervals from Voyager 1, for which the true SF is unknown. As discussed in Section \ref{sec:data}, we take one interval from the inner heliosheath (118au from the Sun), and one from the very local interstellar medium (154au). As mentioned previously, methods of handling gaps in turbulence analysis of Voyager data has largely been limited to improving estimates of the power spectrum \citep{fraternale2017thesis, Gallana2016, Fraternale2019ApJ}. In order to compare our results with these works, after standardizing the intervals and calculating the SFs using \cref{eq:corrected_sf}, we convert \corr\ into an equivalent spectrum using the method developed by \citet{thepthong2024scale}.

\begin{figure*}
    \centering
    \includegraphics[width=7in]{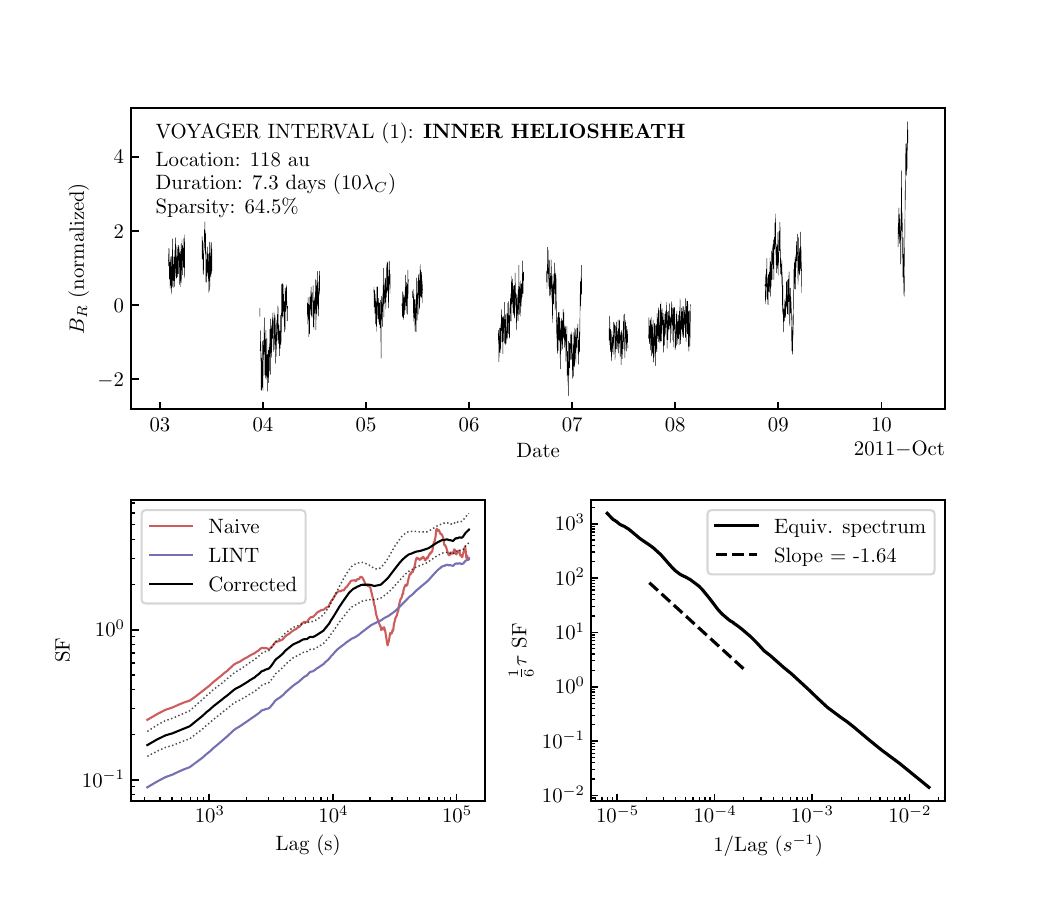}
    \caption{Top: normalized interval of Voyager magnetic field data from the inner heliosheath. Only one of the three vector components used in the calculation is shown, for visualization purposes. Bottom left: \naive, \lint, and \corr\ for the given interval. The dotted black lines indicate \corr\ $\pm$ two standard deviations. Bottom right: equivalent spectrum, calculated from \corr\ according to the procedure given in \citet{thepthong2024scale} (formula given by y-axis label).}
    \label{fig:voyager_ihs}
\end{figure*}

The results for the inner heliosheath interval are given in \cref{fig:voyager_ihs}. \naive\ shows a pronounced trough at around $\tau=3\times 10^4$ s = 8 hours. This is a typical length of contiguous segments in the the inner heliosheath and local interstellar medium, which results in oscillations at multiples of this frequency in both the power spectra and SFs of these regions \citep{Fraternale2019ApJ}.  Interpolation removes the artifact but leads to pronounced decrease in power across the curve, as shown in the LINT estimate (and in \citet[][Supplementary Information, Fig. 1]{Gallana2016}. The corrected estimate, given our validation process, should represent the most accurate curve of the three approaches. From \corr\ we obtain a very smooth and apparently monofractal equivalent spectrum. We calculate the slope in log-log space using least squares regression over the frequency range $2\times10^{-5}\text{s}^{-1}<f<2\times10^{-4}\text{s}^{-1}$. This yields an approximately Kolmogorov inertial range power-law of -1.64. This is similar to the value of -1.72$\pm$0.05 as calculated from the longer but overlapping interval D1 in \citet{Fraternale2019ApJ}.

\begin{figure*}
    \centering
    \includegraphics[width=7in]{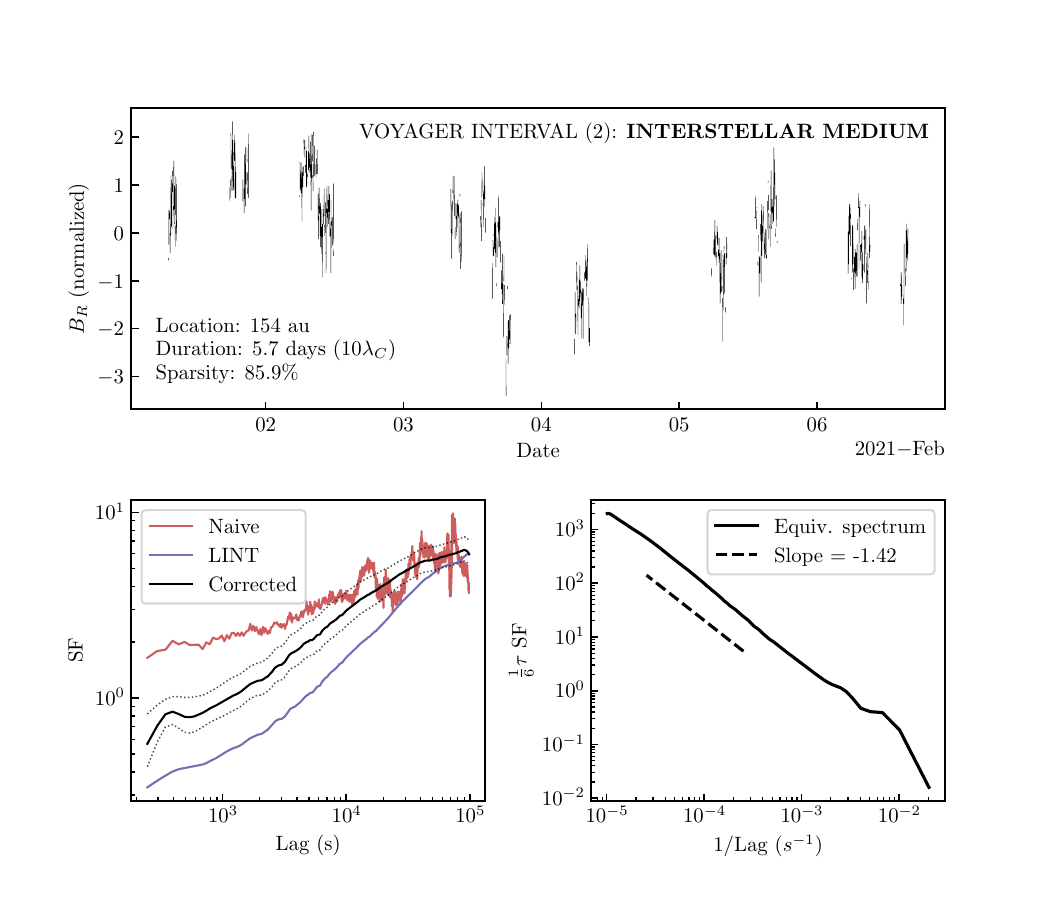}
    \caption{Top: normalized interval of Voyager magnetic field data from the local interstellar medium. Only one of the three vector components used in the calculation is shown, for visualization purposes. Bottom left: \naive, \lint, and \corr\ for the given interval. The dotted black lines indicate \corr\ $\pm$ two standard deviations. Bottom right: equivalent spectrum, calculated from \corr\ according to the procedure given in \citet{thepthong2024scale} (formula given by y-axis label).}
    \label{fig:voyager_lism}
\end{figure*}

The interval from the local interstellar medium in (\cref{fig:voyager_lism}) has more missing data, including both small and large gaps. This introduced to \naive\ both low-frequency oscillations due to periodic gaps, as well as high-frequency noise from individual missing points. These fluctuations are largely absent from \lint\ and consequently the corrected SF. Unlike that in \cref{fig:voyager_ihs}, the equivalent spectrum shows a second, steeper power law at high frequencies, possibly indicating the onset of the dissipation range. In the same low-frequency range used in the previous interval, we obtain a shallower-than-Kolmogorov slope of -1.42. Other works have found similarly shallow slopes in the LISM \citep{Fraternale2019ApJ, fraternale2021, burlaga2020}, which was suggested could be the result of noise or intermittency.

We leave the substantive interpretation of these scaling laws to future work; these examples serve primarily to demonstrate the practical application of our SF correction, which in these cases produce results approximately in line with those derived via the power spectrum and a separate suite of gap-handling approaches in the aforementioned studies. We have re-affirmed that interpolation significantly reduces the spurious oscillations inherited from the gap distribution, and then shown that our technique additionally corrects for the decreased power introduced by interpolation, \textit{given the specific gap distribution}.

\section{Discussion} \label{sec:discuss}

These results demonstrate that, by leveraging a large and diverse set of artificially gapped structure functions, it is possible to empirically learn the biases induced by missing data and thereby substantially reduce their impact. This is in contrast to a prior effort by \citet{Wrench2022} to learn the effect of gaps on structure functions. That work showed that a simple neural network, trained on gapped intervals and the corresponding structure functions from the complete intervals, is effective at capturing large-scale features of the SF, compared to naive and LINT methods. However, the machine learning model underperformed for low-gap percentages and exhibited higher overall error when evaluated using MAPE, the metric used in this study. In comparison, the current method offers a simpler, more interpretable solution that achieves improved accuracy across gap conditions.

Our correction also shares methodological similarities to the ``Optimization" method developed by \citet{Fraternale2019ApJ}. That method also aims to de-bias linear interpolation effects, albeit via a genetic algorithm applied to power spectra, not structure functions. The main disadvantage of that approach is that it is limited to estimating only a few ($<$7) ``control points'' of the spectrum, and is therefore unable to capture localized features of the curve.

A further advantage of our approach is that it avoids reliance on statistical imputation models, such as auto-regressive frameworks \citep{Brown1990, Broersen2006} or Gaussian process-based stochastic interpolators \citep{azari2024stats, Friedrich2020}. These techniques, while useful in providing the uncertainties of predicted values, and even capturing higher-order statistics \citep{lubke2023stochastic}, introduce distributional assumptions, which we explicitly avoid in this study so as not to impose artificial structure on the data. On the topic of higher-order statistics, future directions for this work could involve studying the performance of the correction on higher-order structure functions and kurtosis, as well as cautiously comparing it with these sophisticated interpolation techniques.

In addition to these advantages, there are important limitations to our correction. The variation in turbulence parameters such as Reynolds number, Alfvenicity, and intermittency—which are known to vary with radial distance \citep{Parashar2019, ParasharApJS20, Mondal_2025}—was not fully accounted for. While our data was normalized by correlation length and fluctuation magnitudes, the lack of standardization across these additional parameters likely influenced the algorithm's performance. However, we note that this diversity of turbulence conditions in our training and test sets enhances the generality of our findings.

The standardization we did perform, normalizing intervals to 10 correlation times, represents another limitation. Although this facilitates comparison across spacecraft, it constrains direct applicability of the released correction factor to similarly scaled time series. However, we emphasize that the cadence of the data remains flexible, as lag values are expressed as a fraction of correlation time. This means the actual number of data points per interval may vary, but the correction applies consistently, since all lag values are normalized by correlation time. Also, as noted in the Data Product section below, this 10$\lambda_C$ is a parameter in the code that can be easily changed and the corrections re-evaluated accordingly. Finally, while correlation time estimates from sparse data can be imprecise, their use solely for normalization mitigates the impact of this uncertainty.

\section{Conclusion}\label{sec:concl}

Our understanding of a variety of astrophysical and geophysical processes relies on extracting robust statistics from sparsely sampled time series. An area of research where this is particularly relevant is that of solar wind turbulence in the outer heliosphere and interstellar medium, where precious \textit{in situ} data points are few and far between. In this study we provided updated estimates of \textbf{second-order} SFs and spectral indices for these regions, based on a comprehensive examination of the statistical biases introduced by gaps.

We began by conducting an extensive gap simulation of a large number of solar wind intervals from Parker Solar Probe. In order to produce results that were as general as possible, i.e., not specific to one spacecraft or gap distribution, datasets were standardized and data points removed both randomly and in contiguous chunks. From these simulations, we demonstrated the starkly different effects of ignoring vs. linearly interpolating gaps when calculating SFs, with regards to both the magnitude and direction of errors. As shown in \cref{fig:train_psp_error_trend}, the ``naive'' approach of ignoring the errors, which is commonly thought to be satisfactory, is indeed an unbiased estimator. However, in this context, this simply means that there is no \textit{statistical tendency} to over- or underestimate; the errors are still extremely unpredictable and can frequently be far in excess of $100\%$. 

Linear interpolation, on the other hand, while very effective for small missing fractions, has a clear tendency to lead to underestimates of the SF due to its smoothing effect. This effect is particularly damaging at small lags, and this lag-dependence results in artificial scaling laws, some of the key parameters in turbulence analysis. However, we have shown that this bias is predictable enough that it can, to an extent, \textit{be learned and corrected for}, under the assumptions discussed in Section \ref{sec:discuss}. In a data-driven approach to the problem, we calculated the average estimation error from interpolated intervals as a function of lag, \% missing, and power. This was then used to derive an empirical multiplicative correction factor for the interpolated SF. The improvement in estimation accuracy was proven on a test set from the Wind spacecraft, showing a typical reduction in error of about 50\% compared with ignoring or interpolating gaps. Ultimately, we recommend our correction procedure over these other methods for estimating the second-order SF for datasets with missing fractions of greater than 20\%. 

The success of the learned correction factor on the Wind test set gave us confidence to expect good results for other unseen datasets. Therefore, we applied it to two intervals from Voyager, deriving equivalent spectra and power-laws that are approximately in line with previous results. In conclusion, we have demonstrated a new approach to accessing hitherto unreliable scaling dynamics from sparse solar wind time series (e.g., those in \cref{tab:spacecraft_review}), and, pending evaluation, sparsely-sampled astrophysical and geophysical processes more generally.

\section*{Data product}
We have made the validated correction factor available to the community. We provide the specific values obtained from the PSP data and applied to Wind and Voyager, as well as a notebook demonstrating how to apply it to the Voyager dataset. Noting that these apply specifically to standardized intervals comprising approximately 10 correlation lengths, we also provide the codes used to produce the correction factor, thereby allowing for customized corrections for different length intervals. These are all provided on GitHub\footnote{\texttt{sf\_gap\_analysis} codebase: \url{https://github.com/daniel-wrench/sf_gap_analysis}.} under a 2-Clause BSD License and are archived in Zenodo \citep{wrench_2024_software}. Spacecraft data was downloaded through the NASA CDAWeb interface. 

\section*{Acknowledgments}

The authors thank the PSP, Wind, and Voyager instrument teams for the data and NASA GSFC's Space Physics Data Facility for providing access to it. The authors also wish to acknowledge the use of the Rāpoi high performance computing (HPC) cluster, provided by Victoria University of Wellington; and the New Zealand eScience Infrastructure (NeSI) HPC facilities, consulting support and training services. New Zealand's national facilities are provided by NeSI and funded jointly by NeSI's collaborator institutions and through the Ministry of Business, Innovation \& Employment's Research Infrastructure programme. We also thank the anonymous peer reviewer and the Statistics Editor for their constructive feedback on the manuscript. Finally, we thank Mark Bishop, Jago Edyvean and Federico Fraternale for their helpful comments.

\section*{Author contributions}

DW: Conceptualization, Analysis, Drafting, Editing
TNP: Conceptualization, Supervision, Editing

\section*{Appendix}\label{sec:appendix}

\begin{figure}[ht]
    \centering
    \includegraphics[width=3.5in]{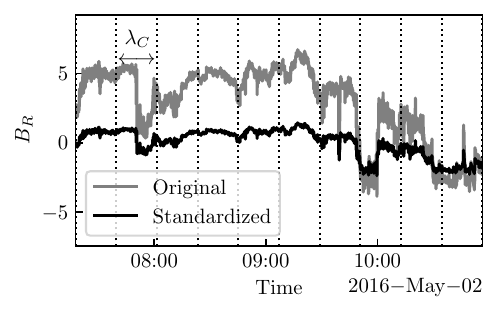}
    \caption{Example of the standardization process for an interval of Wind magnetic field data from 2016. The correlation scale for the entire 24-hour interval was calculated as 22min, using the integral method. The interval was then resampled to 1.3s, to correspond to 10 correlation times across 10,000 points, as indicated by the vertical dotted lines. This allowed for division into two sub-intervals of 10,000 points, the first of which is shown here. Each sub-interval was then standardized to have mean 0 and variance 1, giving the time series in black. (The correlation scale and final intervals used vector data, but only the radial component is shown here for demonstration purposes.)}
    \label{fig:standardise_demo}
\end{figure}

\subsection*{Standardization procedure}

Here we describe the steps of the interval standardization procedure that was outlined in Section \ref{sec:data}. An example of a raw and standardized interval is given in \cref{fig:standardise_demo}.

\begin{enumerate}
    \item Take an interval of magnetic field measurements corresponding to a large number of correlation times, according to typical values for the correlation time from the literature. (For this we use the entire interval covered by each raw file, which conveniently contains about 40-50 correlation times for both PSP and Wind.)
    \item Calculate the \textit{local} correlation time of this entire interval using the integral method \citep{Wrench2024}.
    \item Resample the interval such that 10,000 points corresponds to 10 of these correlation times.
    \item Split the re-sampled interval into sub-intervals of length 10,000 (typically 2-4 of these per original interval). 
    \item If any sub-interval has more than 1\% missing data, discard it. Otherwise, fill in any gaps with linear interpolation, such that each sub-interval is 100\% complete.
    \item Normalize the sub-intervals to have a mean of 0 and variance of 1. 
    \item Calculate the SF from each sub-interval as described in Section \ref{sec:method}. We note that the above procedure results in normalized SFs, as have been used by \citet{Chen_2012}.
\end{enumerate}

% \subsection{3D error heatmaps}

% We can gain further insight into the effect of gaps by also binning the errors based on the power, i.e., the value of the estimated structure function at each lag and \% missing lags. This is ultimately what was used for the correction factor, after smoothing the results with a Gaussian blur. The un-smoothed results are shown in \cref{fig:all_3d_heatmaps}.

% \begin{figure*}
%     \centering
%     \includegraphics[width=6.5in]{all_3d_heatmaps.pdf}
%     \caption{All 3D heatmaps}
%     \label{fig:all_3d_heatmaps}
% \end{figure*}

%=========================================================

\bibliographystyle{aasjournal}
\bibliography{WrenchSFGaps}



\end{document}